\documentclass[aps,twocolumn,graphics,floatfix,tightenlines,amsmath,amsymb,]{revtex4-2}
\usepackage{graphics}
\usepackage{epstopdf}
\usepackage{epsfig}
\usepackage{graphicx}
\usepackage{epsf,epic}
\usepackage{color}
\usepackage{subfig}
\usepackage{amsmath}
\usepackage{booktabs}
\usepackage{multirow}
\usepackage{physics}
\usepackage{hyperref}
\usepackage{amsfonts}
\usepackage{wrapfig}
\usepackage{breqn}
\usepackage{pstricks}
\usepackage{multirow}
\usepackage{fancyref}
\usepackage{pst-node}
\usepackage{float}
\usepackage{bm}
\usepackage{dcolumn}
\newcommand{\etal}{\textit{et al.\ }}
\newcommand{\ie}{\textit{i.e.\ }}

\makeatletter
\usepackage{etoolbox} 
\appto{\appendix}{%
	\@ifstar{\def\theequation@prefix{A.}}%
	{}%
}
\preto\maketitle{%
  \begingroup\lccode`~=`,
  \lowercase{\endgroup
  \let\saved@breqn@active@comma~
  \let~}\active@comma 
}
\appto\maketitle{%
  \begingroup\lccode`~=`,
  \lowercase{\endgroup
  \let~}\saved@breqn@active@comma 
}
\makeatother

\begin{document}
\title{Native defects and their complexes in spinel LiGa$_5$O$_8$: the puzzle of p-type doping} 
\author{Klichchupong Dabsamut}\email{klichchupong.d@gmail.com}
\author{Kaito Takahashi}
\affiliation{Institute of Atomic and Molecular Sciences (IAMS), Academia Sinica, No 1, Sec 4 Roosevelt Road, Taipei, 10617, Taiwan}
\author{Walter R. L. Lambrecht}\email{walter.lambrecht@case.edu}
\affiliation{Department of Physics, Case Western Reserve University, 10900 Euclid Avenue, Cleveland, Ohio 44106-7079, USA}
\begin{abstract}
  Recently, LiGa$_5$O$_8$ was identified as a cubic spinel type ultra-wide-band-gap semiconductor with a gap of about 5.36 eV  and
  reported to be unintentionally p-type. Here we present  first-principles calculations of the native defects and various of their complexes to try to explain
  the occurrence of p-type doping. Although we find Li-vacancies to be
  somewhat shallower acceptors than in LiGaO$_2$, and becoming
  slightly shallower in complexes with donors such
  as V$_{\rm O}$ and Ga$_{\rm Li}$ antisites, these $V_{\rm Li}$ based defects  are
  not sufficiently shallow to explain p-type doping. The dominant defects are donors and
  in equilibrium the Fermi level would be determined by compensation between
  donors  and acceptors, and pinned deep in the gap. 
\end{abstract}
\maketitle
\section{Introduction}
Recently, LiGa$_5$O$_8$  in the spinel structure was grown in thin-film form
by mist-chemical vapor deposition and reported to have a gap of 5.39 eV and
as grown $p$-type conductivity.\cite{Zhao2023} This is a quite surprising finding
as many oxides are notoriously difficult to dope $p$-type. In  $\beta$-Ga$_2$O$_3$
for example, the formation of self-trapped hole polarons related to the heavy valence
band mass is widely believed to preclude $p$-type doping.  Although there are 
some reports of $p$-type doping via complexes, it is not clear whether these
are robust or correspond truly to homogeneous doping and the hole mobilities were found to be rather low.\cite{Chi23,Chikoidze19}
In closely related  LiGaO$_2$, with an even larger optical gap of
6.0 eV\cite{Dadkhah23,Trinkler17,Trinkler22,Tumenas17} native defects were studied \cite{Boonchun19}
and $n$-type doping predicted by Si, and Ge.\cite{Dabsamut20}  However, $p$-type
dopants were not yet identified. Substitutional N$_{\rm O}$ was found to behave amphoteric
and Zn doping suffers from site competition between Zn$_{\rm Li}$ with donor behavior
and Zn$_{\rm Ga}$ acceptor behavior resulting in compensation. Diatomic molecules (N$_2$, NO and O$_2$ in LiGaO$_2$) were also investigated as possible $p$-type dopants, but were all find to
be deep acceptors.\cite{Dabsamut22}  Even in much lower gap ZnO, $p$-type doping has remained notoriously difficult.\cite{Reynolds14}
It would greatly expand the opportunities for power-electronics to have a $p$-type ultra-wide-band gap
UWBG semiconductor but this surprising finding clearly deserves further scrutiny and  requires
first-principles calculations to identify the possible $p$-type dopants.

The spinel structure of LiGa$_5$O$_8$ was established  by Joubert \etal \cite{Joubert63}.
It features Li in octahedral sites and Ga in both octahedral and tetrahedral sites. The compound has received some previous attention as a candidate phosphorescent materials by doping with Cr.\cite{DeClercq17,Sousa20,Huang2018}. Its space group is  $P4_332$ or $O^6$ (No. 212) and its cubic lattice constant is 8.203\AA.\cite{liga5o8struc,Matproj,icsdliga5o8} A figure of the structure, identifying the different types of Ga and O is given in Fig. S1 in Supplemental Material (SM).\cite{supmat}

Recently, one of us \cite{Lambrecht2024} calculated its electronic band structure using the
quasiparticle self-consistent (QS) $G\hat{W}$ method with the screened Coulomb interaction $\hat W$
including ladder diagrams and its optical dielectric function using the Bethe Salpeter Equation (BSE)
method, identifying the quasiparticle gap to be slightly indirect 5.72 eV with lowest direct gap
5.84 eV but a  high exciton binding energy leading to an optical gap of 5.48 eV. These calculations
did not include zero-point motion electron-phonon coupling effects which could reduce the gaps by a few 0.1 eV. To within the error bars this is consistent with the experimental data by
Zhang \etal\cite{Zhao2023}. In \cite{Lambrecht2024} we also showed that Si could provide a shallow $n$-type dopant. 
Here we study point defects using first-principles calculations.  

\section{Computational Methods}
We use the  Vienna Ab initio Simulation Package (VASP)\cite{VASP1,VASP2} using the projector augmented wave (PAW) approach\cite{PAW} and the Heyd-Scuseria-Ernzerhof (HSE) type hybrid exchange correlation functional \cite{HSE03,HSE06} with adjusted inverse screening length $\mu$ and fraction of non-local exchange $\alpha$. We found that $\alpha=0.372$ and $\mu=0.2$ \AA$^{-1}$ gave a direct gap at $\Gamma$ gap of 5.845 eV, and indirect $S-\Gamma$ gap of 5.717 eV, in excellent agreement with the QS$G\hat{W}$ calculations of \cite{Lambrecht2024}.
The band structures are shown for reference in Fig. S2 in SM.\cite{supmat}
We also tested the fulfillment of Koopman's theorem for $V_{\rm Li}$
and Li$_{\rm Ga}$ defects as shown in Fig. S3 in SM.\cite{supmat} We find the cubic lattice constant
$a=8.171$ \AA, slightly smaller than the experimental value of 8.203 \AA.

Because the cubic unit cell (shown in Fig. S1 in SM) already contains 56 atoms or four formula units of LiGa$_5$O$_8$ it is rather challenging to perform calculations for symmetric larger supercells. Even a $2\times2\times2$ cell corresponds to 448 atoms. While for higher convergence, this may be necessary in the future, we here find it more important to study more types of defects and stick to the 56 atom cell.
We use the standard approach of calculating defect energies of formation as
\begin{equation}
	\begin{split}
		\Delta E_{for}(D^q)=E_{tot}(C:D^q)-E_{tot}(C)-\sum_i\Delta n_i \mu_i \\
		+q(\epsilon_v+\epsilon_F+V_{align})+E_{cor},
	\end{split}
\end{equation}
where $D^q$ indicates the defect $D$ in charge state $q$, $C:D^q$ means the crystal with defect
an $C$ without defect, 
$\epsilon_v$ is the valence band maximum (VBM) with respect to the average electrostatic potential which is aligned far away from the defect with that of the perfect crystal by means of the $V_{align}$ correction and $E_{corr}$ is a Madelung correction for the electrostatic energy of the periodic array of localized defect charges in the homogeneous background used to compensate the defect charge.  This and the alignment term for charge defects are obtained using the Freysoldt-Neugebauer-Van de Walle (FNV) procedure\cite{Freysoldt09}. The $\mu_i$ are the chemical potentials of the atoms added to or removed from the perfect crystal go create the defects. 

\section{Results}
\begin{figure}
  \includegraphics[width=8cm]{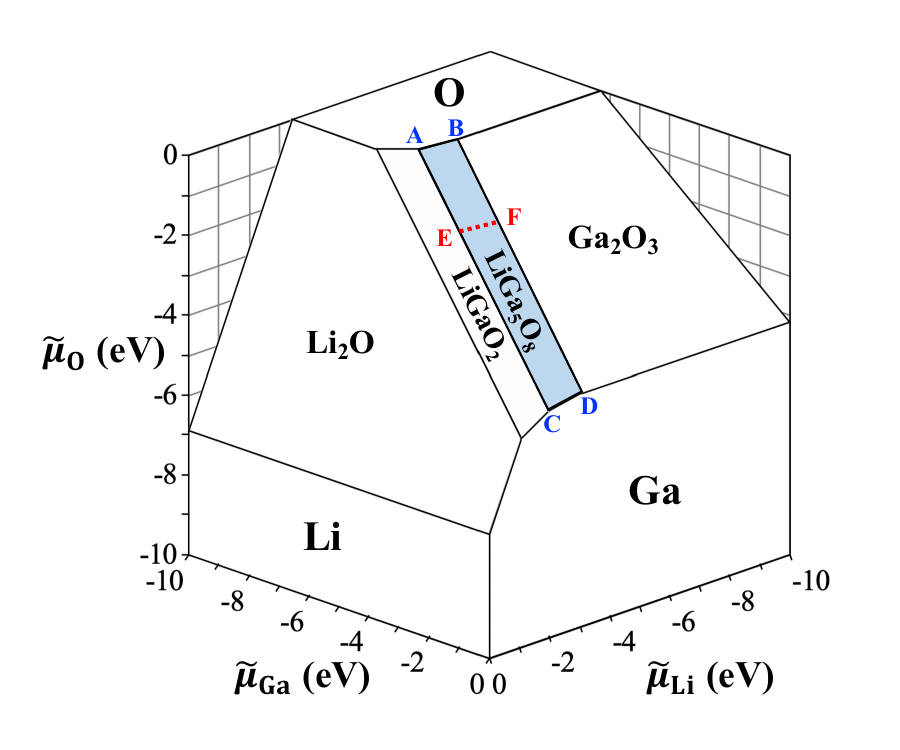}
  \caption{Chemical potential ranges of stability of LiGa$_5$O$_8$. Point E and F correspond to realistic O-chemical potentials at the growth condition. This figure was visualized by the Chesta code.\cite{chesta} \label{fig:chempot}}
\end{figure}

First we determine the chemical potential range in which LiGa$_5$O$_8$ is stable.
As can be seen in Fig. \ref{fig:chempot} this is a rather narrow range. The points  A,B correspond to O-rich conditions (Ga-poor)  and C,D to O-poor (Ga-rich) conditions.
The points E and F correspond to realistic O-chemical potentials corresponding including the pressure and temperature dependent term ($k_BT\ln{p({\rm O}_2)/p_0}$)
\cite{Reuter2001} and at the growth temperature (T=900$^\circ$C) reported in \cite{Zhao2023} and can be seen to be close to O-rich. 
The points A, E and C all correspond to relatively Li rich compared to Ga, while B, F and D are Li-poor.  The actual values of the chemical potentials
at these points is given in Table S1 in SM.\cite{supmat}

\begin{figure}
  \includegraphics[width=8cm]{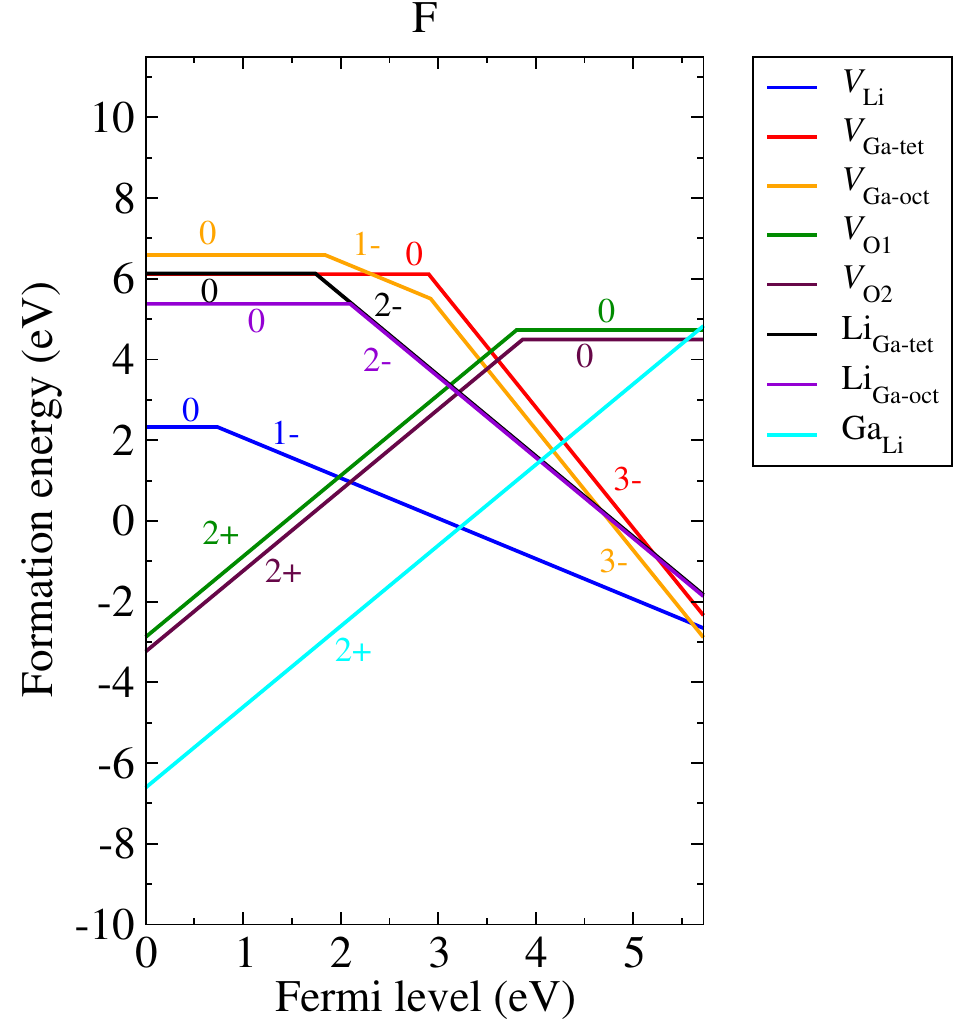}
    \caption{Native defect formation energies for realistic O chemical potential conditions F. \label{figdef}}
    \end{figure}

Our results for simple native point defects are shown in Fig. \ref{figdef}
for the chemical potential conditions F, which corresponds to realistic O-chemical potentials and
somewhat Li-poor. Additional figures for other chemical potential
conditions are given in Fig. S4 in SM. 
We can see that Ga$_{\rm Li}$
acts as a shallow donor, $V_{\rm O}$ are deep donors on both O types.
The $V_{\rm Li}$ is a relatively deep acceptor with acceptor binding energy
or $0/-$ transition level at 0.74 eV. Ga vacancies have higher energy
and the Li$_{\rm Ga}$ acceptors have higher formation energy and higher
$0/-$ transition level  for both the tetrahedral Ga (Ga-tet) and octahedral Ga (Ga-oct) sites.

Consistent with point F being less Li rich, we find (see Fig. S4 in SM\cite{supmat})
that the Li-vacancy
formation energy is increased under condition E compared to F by about 1 eV but Li$_{\rm Ga}$
are lowered in condition E and come closer to the $V_{\rm Li}$. This shifts the
intersection points with Ga$_{\rm Li}$ and $V_{\rm O}$ which will be shown later to be important
but makes no overall qualitative change.  
As also shown in Fig. S4 in SM, under Ga-rich, O-poor conditions (C,D) we find
that  $V_{\rm O}$ defects are lowered in energy by about 4 eV, corresponding to the O chemical potential $(\tilde\mu_{\rm O})$ difference and Ga-vacancies are
even further increased by about 6 eV.  These high-energy of formation defects thus
play little role in the charge neutrality and can safely be ignored.  These Ga-vacancies
can only be expected to be formed under high-energy particle irradiation conditions, which distort
the equilibrium. 

\begin{figure}
  \includegraphics[width=8cm]{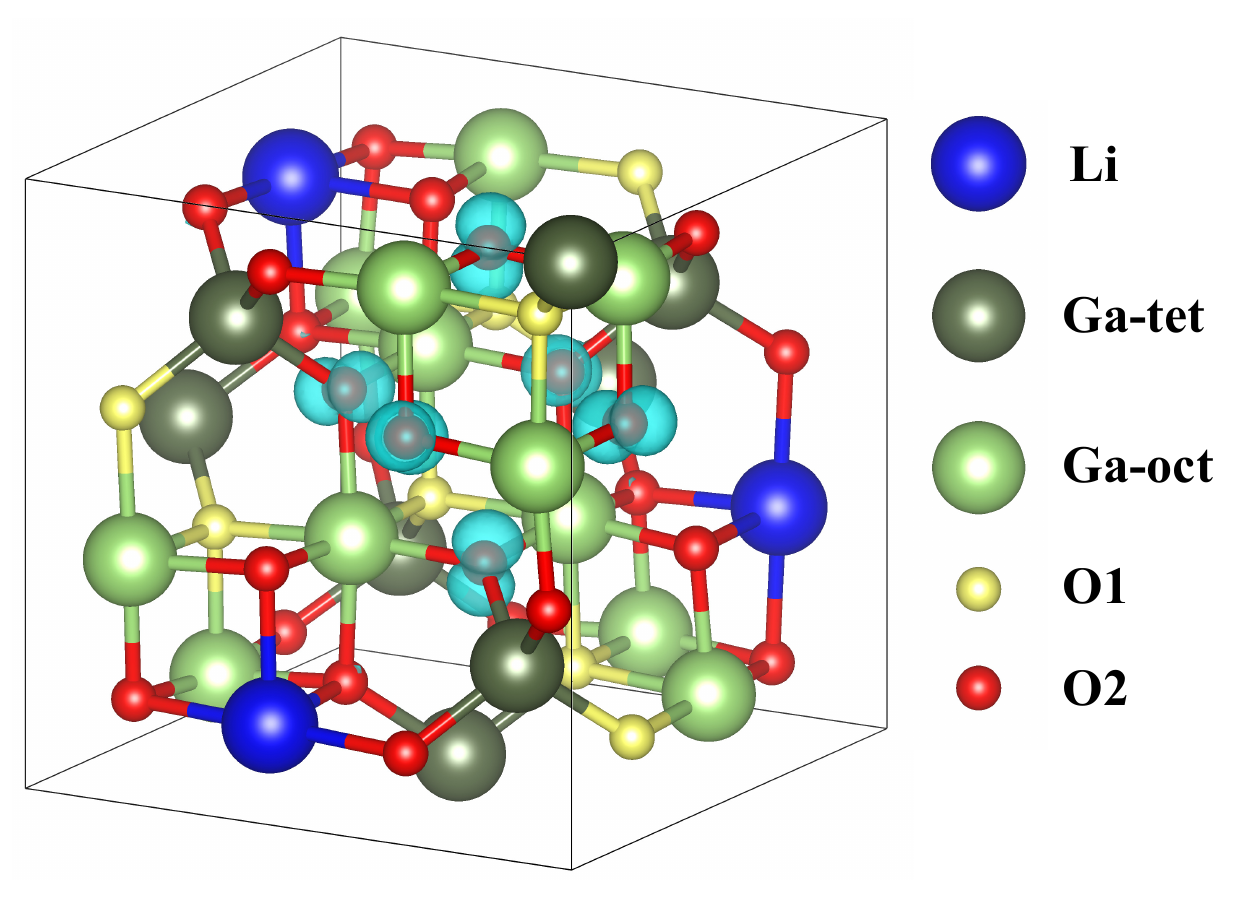}
\caption{$V_{\rm Li}$ wave function (teal colored isosurface ) Dark green spheres: Ga-tetahedral, light green sphere: Ga-octahedral, yellow sphere: O1 which is the O bonded with 4 Ga, red sphere: O2 which is the O bonded with 1 Li and 3 Ga, blue spheres:  Li.\label{fig:vli}}     
\end{figure}

The $V_{\rm Li}$ appears to be the only somewhat viable acceptor but $p$-type doping is not predicted from it because its binding energy is rather deep and 
secondly it would be compensated by O-vacancies and Ga$_{\rm Li}$.
Considering that under conditions E and F the system is rather O-rich and Ga-poor, the Ga$_{\rm Li}$ antisite is apparently a low energy of formation donor.
Nonetheless, it is remarkable that the $0/-$ transition level for the $V_{\rm Li}$
is significantly lower than in LiGaO$_2$, where it is 1.03 eV.\cite{Boonchun19}.
An important difference is that in that material Li occurs in a tetrahedral
environment and its wave function localizes on one of the O neighbors.\cite{Skachkov20} Whereas here it appears to be delocalized over several of the neighboring O atoms as can be seen in Fig. \ref{fig:vli}. An important conclusion is that the $V_{\rm Li}$ does not seem to be a polaronic
acceptor  in the present material. 

\begin{figure}
	\includegraphics[width=8cm]{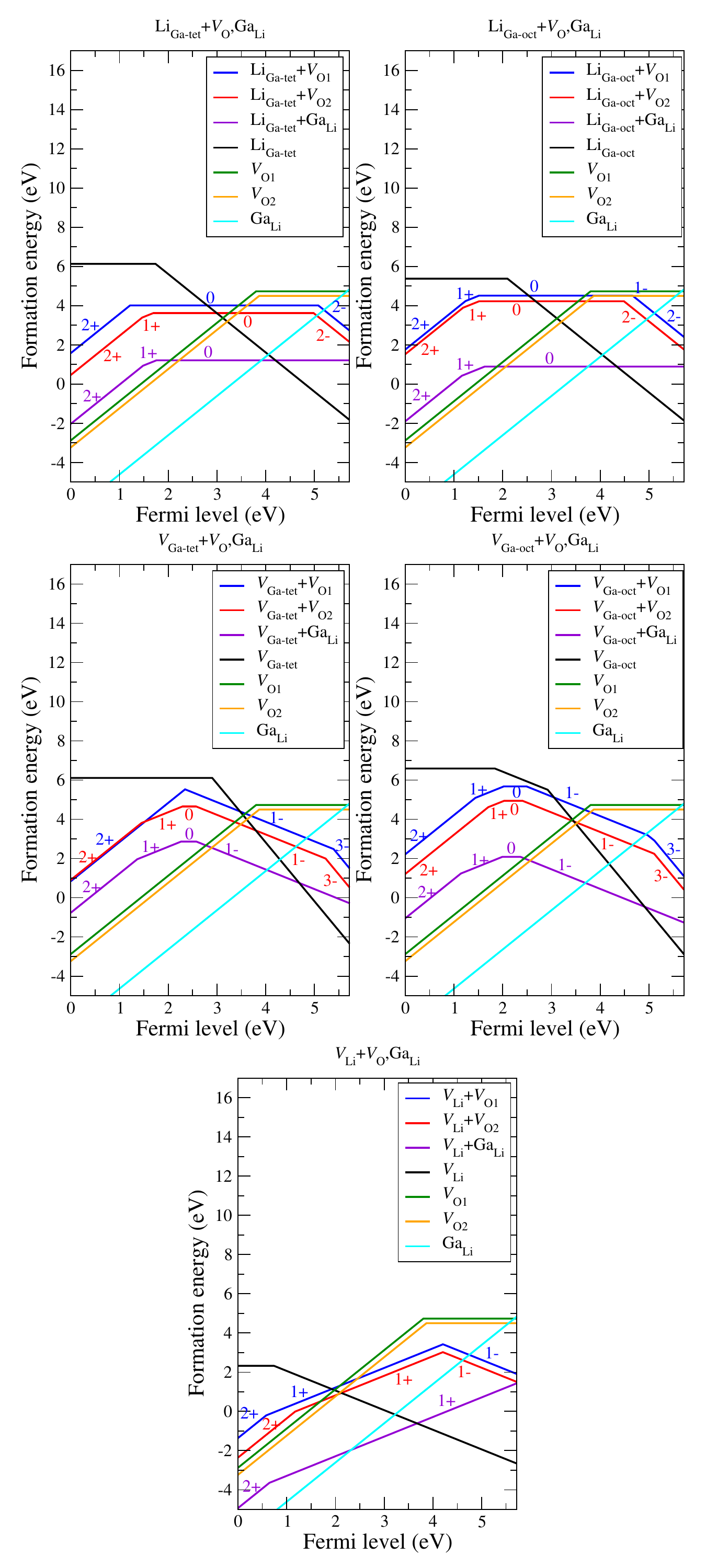}
    \caption{Energies of formation for different charge states for various donor-acceptor pair defect complexes, including Li$_{\rm Ga-tet}$, Li$_{\rm Ga-oct}$, $V_{\rm Ga-tet}$ and $V_{\rm Ga-oct}$ and $V_{\rm Li}$ each paired with three different donors, $V_{{\rm O}_1}$, $V_{{\rm O}_2}$ and Ga$_{\rm Li}$, under chemical potential condition F. \label{defcomplexes}}  
\end{figure}

Next, we study various defect pair complexes. Our reason for doing so is
that we found in our recent study of donor acceptor pairs in LiGaO$_2$ \cite{Dabsamut23}
that the acceptor level of the complex tends to be pushed closer to the VBM and the donor level closer to the CBM. This results from bonding anti-bonding
interactions between their defect wavefunctions. We might thus obtain shallower acceptor levels.
In Fig. \ref{defcomplexes}, we focus again on the chemical potential condition F. A more complete
set of figures for other chemical conditions is given in Fig. S5 in SM.  The subpanels give 
Li$_{\rm Ga-tet}$, Li$_{\rm Ga-oct}$, $V_{\rm Ga-tet}$ and $V_{\rm Ga-oct}$
and $V_{\rm Li}$ each paired with three different
donors, $V_{{\rm O}_1}$, $V_{{\rm O}_2}$ and Ga$_{\rm Li}$ along with the corresponding constitutive point defects. The pairs are chosen as nearest neighbors in each
case.
The relative distance of the defects in the complexes are determined by
Li-O$_1$=3.37 \AA,  Li-O$_2$=2.07 \AA, Li-Li=5.00 \AA,  Li-Ga-tet=3.42 \AA,
Li-Ga-oct=2.84 \AA, Ga-tet-Ga-oct=3.38 \AA, and 
O$_1$-O$_2$=2.66 \AA.   Note that O$_1$ is bonded to 4 Ga while O$_2$ is bonded to a Li and 3 Ga and is thus closer to the $V_{\rm Li}$.

\begin{figure*}
  \includegraphics[width=15cm]{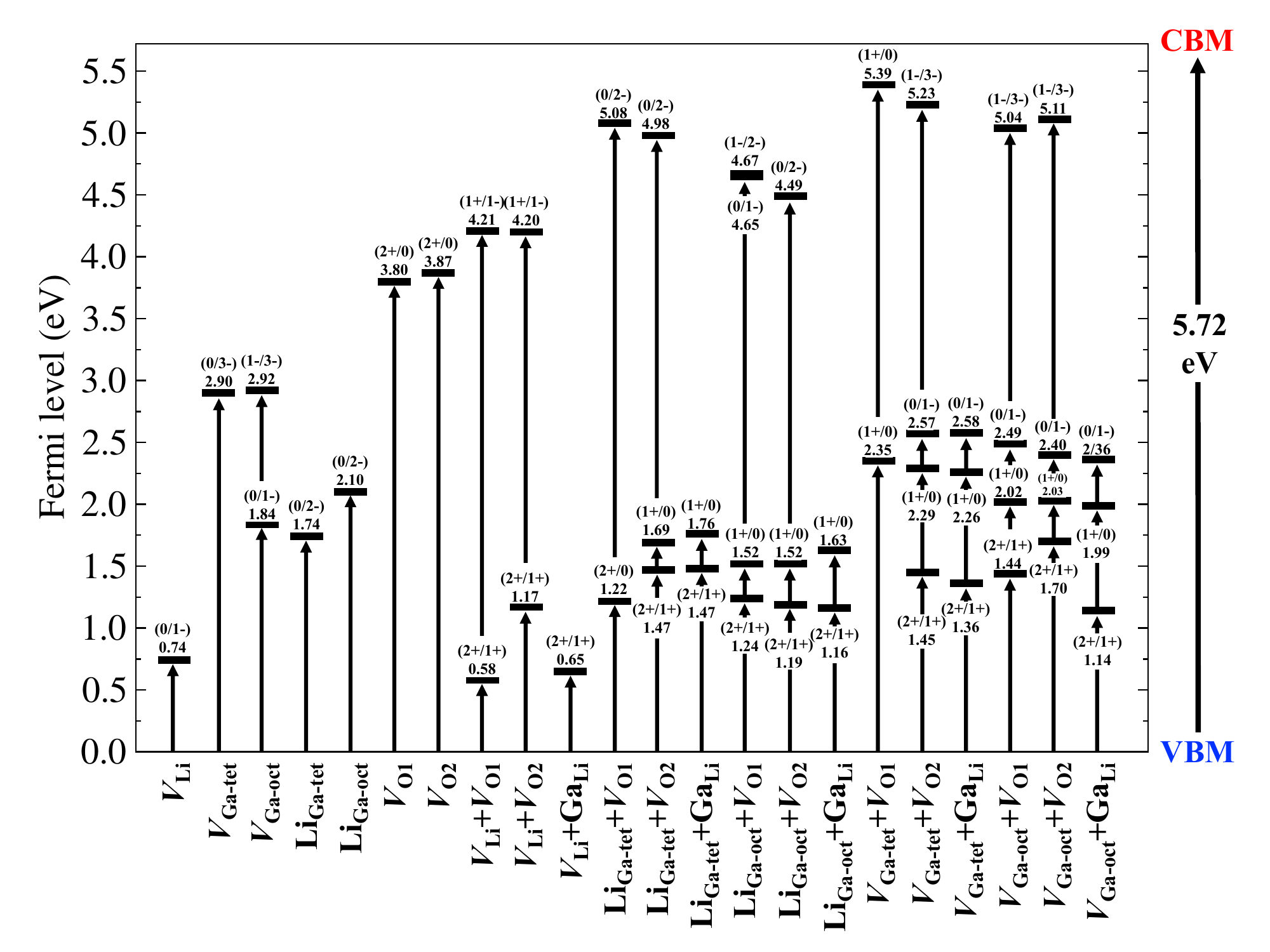}
  \caption{Summary of all transition levels of point defects and their complexes calculated.\label{figlevel}}
\end{figure*}

As  the most relevant  example, let's consider $V_{\rm Li}+V_{{\rm O}_1}$. When the Fermi level
is close to the VBM,  the $V_{{\rm O}_1}$ is in a $2+$ state while the  $V_{\rm Li}$ is neutral.
Increasing the Fermi level, 
the first transition of the complex is $2+/+$.  However, we can see that the $2+/+$ transition level of the complex is indeed closer to the VBM than the $0/-$  level of the isolated $V_{\rm Li}$. Thus, in some sense the acceptor level becomes shallower, and
this corresponds indeed to the $V_{\rm Li}$ emitting a hole to the VBM,
which one might think of as contributing to $p$-type conductivity. 
However, overall the defect is still a donor and is then still in a positive charge state, which requires it to be compensated either by free electrons or by a negative defect. Thus by itself, this complex cannot lead to $p$-type doping
because we combined a double donor with a single acceptor.  Similar considerations apply to  the $V_{\rm Li}$-Ga$_{\rm Li}$ pair. Eventually,
at higher Fermi level position in the gap, we can see a narrow range of 0 charge
state and a transition to the $-1$ charge state. This point again lies
higher than the corresponding donor $2+/0$ transition. 
Unexpectedly, for the $V_{\rm Li}$-$V_{{\rm O}_2}$ complex the $2+/+$ transition is higher than
for the isolated $V_{\rm Li}$.

The Li$_{\rm Ga}$ acceptor is a double acceptor and when it combines with a double donor such as $V_{{\rm O}_1}$, we obtain a $2+/0$, transition and a $0/2-$
transition at high Fermi energy. Thus all the complexes essentially display 
amphoteric behavior, combining donor and acceptor like properties. Close to the VBM, they are dominated by the donor-like behavior and close to the
CBM by the acceptor like behavior. For the double acceptor double donor
pairs a large region of neutral charge state occurs in the center of the gap.
Finally, for the $V_{\rm Ga}$ based complexes, we combine a triple acceptor with
a double donor. However, none of them show promising shallow acceptor like behavior.

An overview of all the transition levels is given in Fig. \ref{figlevel}.
We can see that the  levels closest to the VBM are the $2+/+$ level of the
$V_{\rm Li}+V_{{\rm O}_1}$ and $V_{\rm Li}+{\rm Ga}_{\rm Li}$ at 0.58 eV and 0.65 eV above the VBM.  First of all, these are still not shallow enough to
consider shallow acceptors and secondly, they correspond to $2+/+$ levels,
rather than $0/-$ levels. A shallow level is not sufficient to explain
$p$-type doping. Basically, these are deep donors instead. We also need to consider the overall charge neutrality and compensation issues.  Because the lowest energy defects close to the VBM are the
Ga$_{\rm Li}$ donor and $V_{\rm O}$ type donors, compensation is expected. 

The concentration of defects in various charge states and the net free electron and hole concentrations in the bands are governed by the charge neutrality
condition
\begin{equation}
  -n+p+\sum_i q_i N_i g_i e^{-\Delta E_{for}(D_i^{q_i},\mu)/k_BT},
\end{equation}
where $N_i$ is the density of sites available for a given defect, $q_i$ its charge  and $g_i$
a degeneracy factor. For example if the charged defect features a state with a single electron,
(such as the $V_{\rm Li}^0$) it can be in spin up or spin down state and the degeneracy is 2 but if it is filled or empty, 
the degeneracy is 1. 
Its formation energy in a given charge state depends on the electron chemical potential, 
$\Delta E_{for}(D_i^{q_i})=\Delta E_{for}(D_i^{q_i},\mu=0)+q_i\mu$.
The band electron and hole concentrations $n$ and $p$ also depend on the
electron chemical potential and temperature,
\begin{equation}
  n(\mu,T)=\int_{\epsilon_c}^\infty D(\epsilon)f(\epsilon,\mu,T)d\epsilon
\end{equation}
with $f(\epsilon,\mu,T)=1/(\exp{-(\epsilon-\mu)/k_BT}+1)$ the Fermi distribution.
As long as the chemical potential is not too close to the band edge,
\begin{equation}
  n(\mu,T)=-2\left(\frac{m_n k_BT}{\pi\hbar^2}\right)^{3/2} e^{-(\epsilon_c-\mu)/k_BT}
\end{equation}
with $m_n$ the electron effective mass 
and similar equations hold for the hole concentration.
Together, these equations can be solved for a given temperature,
say the growth temperature, where we assume the system was in
thermodynamic equilibrium before the defects concentrations were frozen in
by quenching the system to room temperature.  They then determine the position
of the chemical potential (\ie the Fermi level) as well as the defect concentrations and free carrier concentrations. 
In practice because of the exponential dependence on the energies of formation
only the lowest energy for formation defects play a significant role and if
the Fermi level stays deep in the gap, the $n$ and $p$ are negligible. 
Keeping only the $V_{\rm Li}^{-1}$, $V_{\rm O}^{+2}$ and for simplicity equating the
formation energy of the two types of O, and Ga$_{\rm Li}^{2+}$
charged defects in the neutrality equation,
and using a growth temperature of $1200$K, we can solve for $\mu$ graphically, using Mathematica and find $\mu=3.25$ eV. This is close to the crossing point of
the $V_{\rm Li}^{-1}$ formation energy with the Ga$_{\rm Li}^{2+}$, indicating
that even the $V_{\rm O}$ may be neglected.  In that case, the equilibrium
Fermi energy is set by
\begin{eqnarray}
  &-&e^{-(\Delta E_{for}(V_{\rm Li}^{-1},\mu=0)+\mu)/k_BT} \nonumber \\
  &+&2e^{-(\Delta E_{for}({\rm Ga}_{\rm Li}^{2+},\mu=0)+2\mu)/k_BT}=0
\end{eqnarray}
because the density of available sites for $V_{\rm Li}$ and Ga$_{\rm Li}$ is the same
and both their degeneracies are 1,
which  can be simplified to $3\mu=kT \ln{2}+\Delta E_{for}(V_{\rm Li})-\Delta E_{for}({\rm Ga}_{\rm Li})$.
Even at $T=1200$ K, $kT\ln{2}=0.072$ eV is pretty small compared to the difference in formation energies of 9.74 eV, but
there is no problem in keeping it and we find $\mu=3.25$ in almost perfect agreement with the calculation including the O-vacancies.  
Thus we conclude that considering only the native defects, the Fermi level is
pinned at the intersection of the dominant donor Ga$_{\rm Li}$ and the
shallowest acceptor-type defect $V_{\rm Li}$. This is under condition F
which is most Li-poor and hence the likeliest to give  Li-vacancies and
Ga on Li antisites.   But the Fermi level is then pinned deep in the gap
and insulating behavior is expected.

Considering now the complexes, these are less likely because the density per unit volume
of $ij$ pairs is $N_iN_j$ and of those only the nearest neighbor ones correspond to the pair considered. In spite of this pre-factor disadvantage of the complex, its energy is lowered by the
bonding between donor and acceptor and as this goes in the exponential, it has a more important
effect. 
Numerical solution of the neutrality equation, shows that $\mu$ now shifts to 3.54 eV, in other words, it shifts close
to the intersection of the $V_{\rm Li}^{-1}$ and the $[V_{\rm Li}-{\rm Ga}_{\rm Li}]^{+1}$, which now becomes the lowest energy
donor. So, even though the $2+/+$ level moved closer to the VBM, the lower energy of these complexes ultimately shifts the Fermi level
even higher in the gap. 

We also consider what happens to the chemical potential at room temperature
if we assume the defect concentrations from  the growth temperature are frozen
in but their charge state might change and including the free carrier
concentrations.  When we only include the native defects, we find that the  chemical potential then drops to
0.82 eV above the VBM but the hole concentration is still negligible ($~\sim10^6$ cm$^{-3}$)
and the neutrality is still produced by the equilibrium between Ga$_{\rm Li}^{2+}$
and $V_{\rm Li}^{-1}$  defects which have a concentration
of 2.1$\times10^{22}$ cm$^{-3}$ and 4.2$\times10^{22}$ cm$^{-3}$ respectively.
We indeed need twice the concentration of single negative acceptors as
2+ donors to obtain neutrality. When we include the lowest energy of formation defect pairs, we
find the room temperature chemical potential at 0.83 eV, the hole and electron concentrations are negligible (p$\approx1.4\times10^6$ cm$^{-3}$)
and the neutrality  then comes from the balance of $V_{\rm Li}^{-1}$ with $[V_{\rm Li}-{\rm Ga}_{\rm Li}]^{+1}$
complexes, both of which are found around 7.2$\times10^{23}$ cm$^{-3}$ with a smaller concentration of Ga$_{\rm Li}^{+2}$ of or 7$\times10^{19}$ cm$^{-3}$. 

Thus in thermal equilibrium, even if we include complexes, our calculations
unambiguously predict insulating behavior because the Fermi level stays too far away from the
valence band edge. 

\begin{figure}
  \includegraphics[width=7cm]{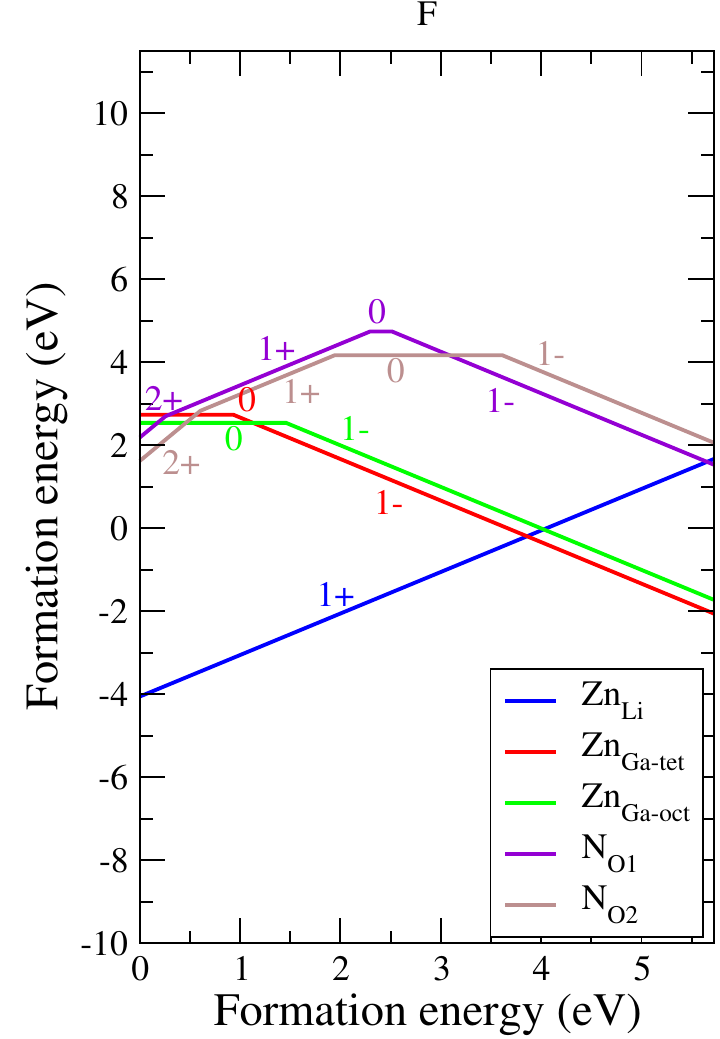}
  \caption{Energies of formation for N and Zn dopants.\label{figdopants}}
\end{figure}

We also considered some impurity doping candidates for $p$-type doping such
as N substituting for O and Zn substituting for Ga. The results are shown
in Fig. \ref{figdopants}. We can see that substitutional N on both O$_1$ or O$_2$
sites behaves as an amphoteric defect. Although its lowest level is fairly
close to the VBM, it is a $2+/+$ transition, typical of a deep donor.
For Zn$_{\rm Ga}$, we find acceptor behavior for both tetrahedral and octahedral site with slightly shallower transition level $0/-$ for the tetrahedral site
but both are about 1 eV or larger above the VBM. Furthermore, there is
site competition with Zn$_{\rm Li}$ which acts as a shallow donor, in the $+1$
charge state throughout the gap.

\section{Discussion and Conclusion}
In spite of  a rather exhaustive study of the possibilities, our calculations
are unable to explain $p$-type doping in LiGa$_5$O$_8$. 
In equilibrium we clearly find donors like Ga$_{\rm Li}$ to compensate
the potential $V_{\rm Li}$ acceptor. The Li vacancy acceptor itself is somewhat shallower
than in LiGaO$_2$ where it occurs in tetrahedral coordination
as opposed to octahedral coordination found here. This different environment
does avoid the polaronic distortion and we find a more spread out vacancy wave function,
but, despite this being somewhat encouraging,  the level is still too deep in the gap
to provide $p$-type behavior.
Complex formation with $V_{\rm O}$  or with Ga$_{\rm Li}$ can push the lowest transition levels closer to the VBM.
Such complexes have less available sites but may nonetheless become slightly favored by the binding energy 
lowering of the energy of formation. However, these complexes are net amphoteric or donors rather than acceptors.
Thus, they play the role of compensating the isolated $V_{\rm Li}$ acceptors but
do not become acceptors themselves.  The  lowest energy  complex found here is
the $V_{\rm Li}$-Ga$_{\rm Li}$ pair and because it is lower in energy and becomes a single donor
it shifts the Fermi level even deeper in the gap. Even if we consider that,  after cooling
to room temperature, the defects may change their charge state and thereby affect the free carrier concentrations, 
we find the Fermi level to still be at about 0.8 eV  above the VBM, which is too high to give $p$-type doping.
The maximum $p$-type concentration we obtain in this way is of order $10^6$ cm$^{-3}$.

The only way to explain the observation of $p$-type conduction is then
a non-equilibrium situation. One might envision, regions of the sample
where $V_{\rm Li}$ occur  or their complexes occur and emit holes but these
$p$-type regions would then have to be compensated by opposite charge
space charge regions. This non-equilibrium could possibly occur near
grain boundaries, or other extended defects or near surfaces.

In \cite{Zhao2023} $p$-type doping is reported for various growth conditions
and in that sense labeled robust. Notably they report it to occur for both Li-rich
and Li-poor conditions. However, as we noted the Li-chemical potential range in which
LiGa$_5$O$_8$ is stable is rather narrow, so this is perhaps not too surprising.
These authors also report that annealing in O$_2$  could render samples which were initially
$p$-type to become insulating.  This might lead one to  the assumption that oxygen vacancies play
a role in the $p$-type conductivity. However, O-vacancies are clearly donors and not acceptors. Even if they could play a role in complex formation with $V_{\rm Li}$ and thereby
push the $2+/+$ transition of the complex closer to the VBM  than the $0/-$ transition
of the acceptor, this in itself cannot explain $p$-type doping. 
The O-annealing effects could perhaps rather point to a surface region annealing effect
and restore the native insulating behavior we predict from our defect calculations.
Although our results at present did not find
any plausible explanation for $p$-type doping, we hope that the results on defect levels in this system will prove useful to its further characterization.




\acknowledgements{This work was supported by the US. Air Force Research Office (AFOSR)
  under grant No. FA9550-22-1-0201. It made use of the High Performance Computing Resource in the Core Facility for
  Advanced Research Computing at Case Western Reserve University as well as additional resources at IAMS.
  K.D. acknowledges support by Academia Sinica.
}

 \bibliography{ga2o3,ligao2,liga5o8,dft,defects}
\end{document}